\documentclass[pre,aps,twocolumn,showpacs]{revtex4}
\usepackage{graphics}
\usepackage{epsfig}

\begin{document}

\title{Mapping of 2+1-dimensional Kardar-Parisi-Zhang growth onto a
    driven lattice gas model of dimers}

\author{G\'eza \'Odor (1), Bartosz Liedke (2) and Karl-Heinz Heinig (2)}

\affiliation{(1) Research Institute for Technical Physics and
  Materials Science, \\
P.O.Box 49, H-1525 Budapest, Hungary\\
(2) Institute of Ion Beam Physics and Materials Research \\
Forschungszentrum Dresden - Rossendorf \\
P.O.Box 51 01 19, 01314 Dresden, Germany}    

\begin{abstract}
We show that a $2+1$ dimensional discrete surface growth model 
exhibiting KPZ class scaling can be mapped onto a two dimensional 
conserved lattice gas model of directed dimers. In case of KPZ 
height anisotropy the dimers follow driven diffusive motion. 
We confirm by numerical simulations that the scaling exponents 
of the dimer model are in agreement with those of the $2+1$
dimensional KPZ class. This opens up the possibility of 
analyzing growth models via reaction-diffusion models, 
which allow much more efficient computer simulations. 
\end{abstract}
\maketitle

\section{Introduction}

The Kardar-Parisi-Zhang (KPZ) equation \cite{KPZeq} motivated by
experimentally observed kinetic roughening has been the subject 
of large number of theoretical studies \cite{HZ95,krug-rev}. 
Later it was found to model other important physical phenomena 
such as randomly stirred fluid, \cite{forster77}, dissipative 
transport \cite{beijeren85,janssen86}, directed polymers
\cite{kardar87} and the magnetic flux lines in superconductors \cite{hwa92}.
It is a non-linear stochastic differential equation, which describes
the dynamics of growth processes in the thermodynamic limit specified
by the height function $h({\bf x},t)$
\begin{equation}
\label{KPZ-e}
\partial_t h({\bf x},t) = v + \sigma\nabla^2 h({\bf x},t) + 
\lambda(\nabla h({\bf x},t))^2 + \eta({\bf x},t) \ .
\end{equation}
Here $v$ and $\lambda$ are the amplitudes of the mean 
and local growth velocity, 
$\sigma$ is a smoothing surface tension coefficient 
and $\eta$ roughens the surface by a zero-average Gaussian 
noise field exhibiting the variance
\begin{equation}
\langle\eta({\bf x},t)\eta({\bf x'},t')\rangle 
= 2 D \delta^d ({\bf x-x'})(t-t') \ .
\end{equation}
Here $d$ is used for the dimension of the surface, $D$ for the 
noise amplitude and $\langle\rangle$ denotes average 
over the noise distribution.
In $1+1$ dimensions it is exactly solvable \cite{kardar87},
but in higher dimensions only approximations are available
(see \cite{barabasi}). In $d>1$ spatial dimensions due to the
competition of roughening and smoothing terms, 
models described by the KPZ equation exhibit a roughening 
phase transition between a weak-coupling regime ($\lambda<\lambda_c$), 
governed by $\lambda=0$ Edwards-Wilkinson fixed point \cite{EWc}, 
and a strong coupling phase.
The strong coupling fixed point is inaccessible by perturbative 
renormalization method. Therefore the KPZ phase space has been
the subject of controversies and the value of the upper critical 
dimension has been debated for a long time.
Very recently non-perturbative renormalization and mode coupling theory
has revealed a rich phase diagram, with more than one lines of fixed point
solutions in the $d$ - $\lambda$ space \cite{CM07}. This suggests an
upper critical dimension $d_c=4$ for KPZ, but an earlier numerical work 
\cite{MPPR02} does not support this claim.

\begin{figure}[ht]
\begin{center}
\epsfxsize=75mm
\epsffile{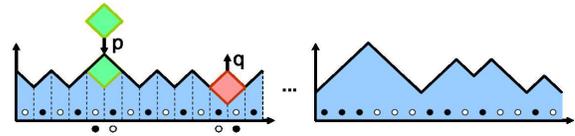}
\caption{(Color online) Mapping of the $1+1$ dimensional
surface growth onto the $1d$ ASEP model. 
Surface attachment (with probability $p$) and detachment 
(with probability $q$) corresponds to anisotropic diffusion 
of particles (bullets) along the $1d$ base space.} \label{kpzasep}
\end{center}
\end{figure}

In one dimension a discrete, restricted solid on solid
realization of the KPZ growth is equivalent to the asymmetric 
simple exclusion process (ASEP) of particles \cite{kpz-asepmap,meakin} 
(see Fig.~\ref{kpzasep}).

In this discrete so-called 'roof-top' model the heights are quantized 
and the local derivatives can take the values $\Delta h = \pm 1$. 
By considering the up derivatives ($\Delta h=1$) as particles and 
down ones as holes the roughening dynamics can be mapped onto 
a driven diffusive system of particles with single site occupancy.
The ASEP model on the other hand is well known and its scaling 
properties are explored (for a recent review see \cite{EH05}). 

The extension of this kind of lattice-gas analogy to higher 
dimensions has not been considered to our knowledge. Instead
hypercube stacking models were constructed \cite{meakin,FT90L,FT90J}
and surface configurations were mapped onto the $d$-state
Potts spins defined on the substrate lattice itself. 
Especially $2+1$ dimensional surfaces were shown to be related 
to the six-vertex model with equal vertex energies \cite{Baxter} 
and to the ground-state configurations of the anisotropic Ising model 
defined on the triangular lattice \cite{BH82}. 
As a consequence the height-height correlation functions can be 
related to four-spin-correlation functions of the spin system.
Very recently the conformal invariance of the isoheight lines
has also been pointed out \cite{sab08}. 

Here we show that a $2+1$ dimensional growth model exhibiting KPZ
scaling can also be mapped onto a driven lattice gas.
This is important from theoretical point of view, because the
scaling behavior of driven diffusive system (DDS) has been studied
intensively for a long time (for a review see ref.~\cite{S-Z}),
thus results for DDS may be exploited to understand KPZ better 
and vica versa \cite{TF02}. Furthermore this conserved lattice 
gas and its generalizations with anisotropies, disorder, or 
higher order terms can be studied effectively by bit-coded 
algorithms for example.

\section{Mapping onto lattice gas in two dimensions}

\begin{figure}[ht]
\begin{center}
\epsfxsize=70mm
\epsffile{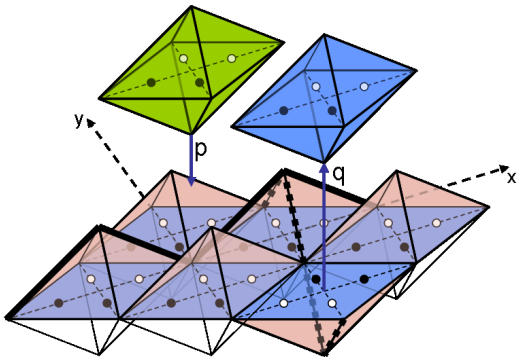}
\caption{(Color online) Mapping of the $2+1$ dimensional surface 
growth onto the $2d$ particle model (bullets). 
Surface attachment (with probability $p$) and
detachment  (with probability $q$) corresponds to 
Kawasaki exchanges of particles, or to anisotropic 
diffusion of dimers in the bisectrix direction 
of the $x$ and $y$ axes. The crossing points of dashed lines 
show the base sub-lattice to be updated. Thick solid/dashed
lines on the surface show the $x$/$y$ cross-sections, 
corresponding to the $1d$ model (Fig.~\ref{kpzasep}.)} 
\label{2dModel}
\end{center}
\end{figure}

As a generalization of the $1+1$ dimensional roof-top model,
where the building blocks are squares let's put octahedra on
the square lattice, such that we get back the $1+1$ dimensional 
model in the $x$ or $y$ direction as shown on Figure \ref{2dModel}.
Surface adsorption or desorption events correspond to
attachment or detachment of octahedra, respectively.
The surface built up from the octahedra can be described by the
edges meeting in the up/down middle vertexes. The up edges in the 
$x$ or $y$ directions are represented by '$+1$'-s, while the 
down ones by '$-1$' in the model. In this way a single site 
deposition flips the four edges and means two 
'$+1$'$\leftrightarrow$'$-1$' (Kawasaki) exchanges: 
one in the $x$ and one in the $y$ direction. This can also be 
understood as a special $2d$ cellular automaton \cite{Wolfram}
with the generalized Kawasaki updating rules
\begin{equation}\label{rule}
\left(
\begin{array}{cc}
   -1 & 1 \\
   -1 & 1 
\end{array}
\right)
\rightleftharpoons
\left(
\begin{array}{cc}
   1 & -1 \\
   1 & -1 
\end{array}
\right)
\end{equation}
with probability $p$ for attachment and probability $q$ for
detachment.
We can also call the '$+1$'-s as particles and the '$-1$-s as holes of 
the base square lattice. In this way an attachment/detachment update
can be mapped onto a single step motion of an oriented dimer in the
bisectrix direction of the $x$ and $y$ axes. To make a one-to-one 
mapping we update the neighborhood of sub-lattice points, which are 
denoted by the crossing-points of the dashed lines only.

Since the three dimensional space can't be filled fully by octahedra,
holes can occur among them, below the surface.  Therefore this 
approximation of a surface growth may not sound to be faithful 
and the validity of KPZ growth rules requires confirmation.
Note, however that in reality atoms are not cubes either 
and do not tile the three dimensional space completely.
Furthermore very recently in bi-disperse ballistic deposition models 
\cite{bid1,bid2}, in which under-surface vacancies may occur KPZ scaling  
has been reported as well.

The deterministic part of the KPZ equation (\ref{KPZ-e}),
which can be obtained by averaging over the noise can be 
derived from the surface/dimer model similarly as it was done in 
$1+1$ dimension \cite{kpz-asepmap}. If we apply the transformation
\begin{equation}\label{BKmap}
{\bf v}({\bf x},t) = \nabla  h({\bf x},t)
\end{equation}
we get the Burgers equation for the height profile 
\begin{equation}\label{BUR-e}
\partial_t {\bf v}({\bf x},t) = \sigma\nabla^2 {\bf v}({\bf x},t) + 
\lambda {\bf v}({\bf x},t) \nabla  {\bf v}({\bf x},t) \ .
\end{equation}

Our system is represented by two matrices $\Delta _{x}$ and $\Delta
_{y}$\ of sizes $L\times L$, which contain discrete derivatives 
$+1$ or $-1$ in $x$ and $y$ direction, respectively 
(see Eqs.~(\ref{up1}),(\ref{up2})).
In two dimensions we introduce the vector variable $\overline\sigma_{i,j} =
(\Delta_x(i-1,j),\Delta_y(i,j-1))$. This has the value $(1,1)$
in case of a dimer and $(-1,-1)$ for a pair of holes.
By setting up the master equation 
\begin{eqnarray}
\partial_t P( \{\overline\sigma\},t) &=&\sum_{i,j} 
w'_{i,j}(\{\sigma\}) P( \{\overline\sigma'\},t) \nonumber \\
&-& \sum_{i,j} w_{i,j}(\{\overline\sigma\}) P( \{\overline\sigma\},t) 
\end{eqnarray}
for the probability distribution $P(\{\overline\sigma\},t)$,
where the prime index denotes a state as a result of a 
generalized Kawasaki flip (\ref{rule}) the transition 
probability is given by
\begin{eqnarray} \label{mast}
w_{i,j}(\{\overline\sigma\}) &=&
\frac{1}{8} [ 2 - \overline\sigma_{i+1,j+1}\overline\sigma_{i,j} 
+ \lambda (\overline\sigma_{i+1,j+1} - \overline\sigma_{i,j}) \nonumber \\
&-& \frac{(1-\lambda)}{2} 
(\overline\sigma_{i+1,j+1}\times\overline\sigma_{i,j})^2 ] \ ,
\end{eqnarray}
with $\lambda = 2\frac{p}{p+q}-1 $ parametrization. 
This formally looks like the one-dimensional Kawasaki exchange 
probability (shown in \cite{kpz-asepmap}), except the cross-product term, 
which is necessary to avoid surface discontinuity creation.
The cross-product as a determinant cancels updates between
configurations like $(1,1) \to (1,-1)$. The nonlinear term
vanishes for $p=q$ ($\lambda=0$). The sign of the coefficient $\lambda$
of the nonlinear term can be interpreted as follows. For $p>q$ 
positive nonlinearity (positive excess velocity) it is a 
consequence of growth with voids.

To obtain Eq.~(\ref{BUR-e}) first one averages over the slope vectors
\begin{equation}
\langle\overline\sigma_{i,j}\rangle =\sum_{\{\overline\sigma\} }
\overline\sigma_{i,j}P( \{\overline\sigma\},t) \ .
\end{equation}
Then calculating its time derivative using the master equation 
the cross-product term drops out and one obtains
\begin{eqnarray}
2 \partial_t\langle \overline\sigma_{i,j} \rangle &=&
\langle \overline\sigma_{i-1,j-1}\rangle - 
2\langle\overline\sigma_{i,j} \rangle +
\langle \overline\sigma_{i+1,j+1}\rangle \nonumber \\
&+& \lambda\langle\overline\sigma_{i,j}(\overline\sigma_{i+1,j+1} -
\overline\sigma_{i-1,j-1})\rangle \ .
\end{eqnarray}
Here one can see the discrete second and first differentials of 
$\overline\sigma_{i,j}$ corresponding to the operators of (\ref{BUR-e}).
These differentials are one-dimensional because the dimer dynamics
is also one-dimensional.
Making a continuum limit in both directions and taking into account 
the relation of height and slope variables (\ref{BKmap}) we can 
arrive to the deterministic part of the KPZ equation (\ref{KPZ-e}). 

This agreement does not prove the equivalence of
KPZ and the dimer model since they are just the first equations 
in the hierarchy of equations for correlation functions. 
On the other hand from universal scaling point of view they 
show the equivalence of the leading order terms.
We will show by numerical simulation that our mapping is faithful
and reproduces the KPZ class surface growth behavior. 

\section{The simulation algorithm}

In the algorithm we extend the sequence of discrete slopes of the $1$d 
ASEP model (Fig.~\ref{kpzasep}) to local derivatives at $(i,j)$ sites 
in $x$ and $y$ directions of the surface (see Fig.~\ref{2dModel}). 
The initially flat surface is presented as a regular sequence 
of '$+1$'-s and '$-1$'-s within both matrices. Periodic boundary conditions
are applied to $x$ and $y$ direction. The system's evolution is simulated as
follows. 

A site $(i,j)$ on the substrate plane is selected randomly. Then, we choose
an attachment or detachment attempt according to their probabilities $p$ and 
$q$. Generalized Kawasaki exchanges (\ref{rule}) 
of attachment or detachment are realized if 
\begin{eqnarray} \label{up1}
\left( 
\begin{array}{cc}
\Delta _{x}(i-1,j) & \Delta _{x}(i,j) \\ 
\Delta _{y}(i,j-1) & \Delta _{y}(i,j)%
\end{array}%
\right) &=& \left( 
\begin{array}{cc}
-1 & 1 \\ 
-1 & 1%
\end{array}%
\right) \\ 
{\rm or} \ \ 
\left( 
\begin{array}{cc}
\Delta _{x}(i-1,j) & \Delta _{x}(i,j) \\ 
\Delta _{y}(i,j-1) & \Delta _{y}(i,j)%
\end{array}%
\right) &=& \left( 
\begin{array}{cc} \label{up2}
1 & -1 \\ 
1 & -1%
\end{array}%
\right) \ ,
\end{eqnarray}
respectively.
Throughout this paper the time is measured by Monte Carlo steps (MCS), 
i.e. $L\times L$ jump attempts correspond to one MCS. 
After certain time intervals data evaluation requires the reconstructions of
the surface heights $h_{x,y}\left( t\right) $ by summing up the sequence of
local slopes $\Delta_x$, $\Delta_y$. 

\section{Results}

Starting from periodic, vertically striped particle distribution,
which corresponds to a flat initial surface we update the particle
model by the rules defined in the previous section. 
At certain time steps we calculate the $h_{x,y}(t)$ heights from the  
height differences $\Delta_{x,y}$.
The morphology of a growing surface is usually characterized
by its width
\begin{equation}
W(L,t) =
\Bigl[
\frac{1}{L^2} \, \sum_{x,y}^L \,h^2_{x,y}(t)  -
\Bigl(\frac{1}{L^2} \, \sum_{x,y}^L \,h_{x,y}(t) \Bigr)^2
\Bigr]^{1/2} \ .
\end{equation}
In the absence of any characteristic length, growth processes are expected to
show power-law behavior of the correlation functions in space and height and
the surface is described by the Family-Vicsek scaling~\cite{family}
\begin{eqnarray}
\label{FV-forf}
W(L,t) &\propto& t^{\beta} , \ \ {\rm for} \ \ t_0 << t << t_s \\
       &\propto& L^{\alpha} , \ \ {\rm for} \ \ t >> t_s \ . \label{FV-a}
\end{eqnarray}
Here $\alpha$ is the roughness exponent and characterizes 
the deviation from a flat surface in the stationary regime 
($t>>t_s$), in which the correlation length has exceeded 
the linear system size $L$; 
and $\beta$ is the surface growth exponent, which describes the 
time evolution for earlier (non-microscopic $t>>t_0$) times.
The dynamical exponent $z$ can be expressed by the ratio
\begin{equation}\label{zlaw}
z = \alpha/\beta \ .
\end{equation}

\begin{figure}[ht]
\begin{center}
\epsfxsize=70mm
\epsffile{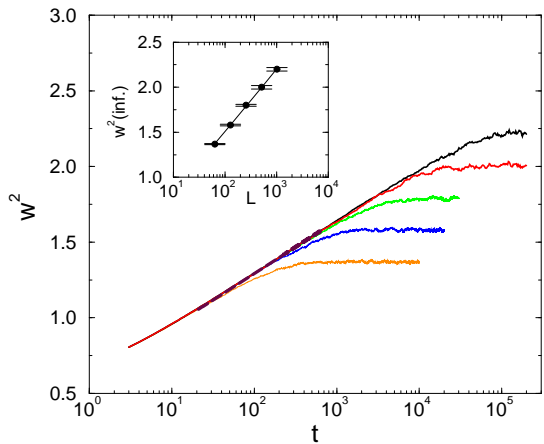}
\caption{(Color online) Logarithmic surface growth in case of
up-down symmetry for different sizes $L=64,128,256,512,1024$
(bottom to top). The dashed line shows the fitting with the
form (\ref{logform}). Inset: width saturation values for different 
system sizes $L$ in the long time limit.} \label{EWcoll}
\end{center}
\end{figure}

In case of up-down symmetry ($p=1$, $q=1$) the nonlinear term
is dropped, and the KPZ equation (\ref{KPZ-e}) 
simplifies to the Edwards-Wilkinson (EW) equation \cite{EWc}. 
Since the upper critical dimension of this equation is: $d_c=2$,
mean-field behavior, characterized by $\alpha=\beta=0$ and
logarithmic scaling is expected by field theory.
Indeed, the width of the surface grows like
\begin{equation}\label{logform}
W^2(t) = a \ln(t) + b
\end{equation}
as shown in Fig.~\ref{EWcoll}.
The prefactor $a$ obtained by fitting the $L=1024$ curve
in the $20 < t < 1000$ region with the form (\ref{logform}) is
$a=0.152(8)$. This is in agreement with the theoretical estimate
for the EW equation $D/(4\pi\sigma)$ \cite{NatTang92}  
if take into account the exact value for the stiffness
constant (or surface tension): $\bar\sigma/D=\pi/9$.
This constant was identified by \cite{BH82} through the 
correspondence between the exact calculation of the four-spin 
correlation function of the zero-temperature triangular Ising 
antiferromagnet \cite{Step64} and the discrete height-height 
correlation function in real space in the interface model.
A factor $\bar\sigma=2/3\sigma$ is coming from the $2/\sqrt 3$
triangular lattice site per surface element and the $1/\sqrt 3$ of the 
octahedron/cube surface fraction,
thus the theoretical estimate is: $a \simeq 0.151981$.

The saturation values are expected to exhibit logarithmic growth
\begin{equation}
W^2(inf.) = \lim_{t\to\infty} W^2(t) = c\ln(L)+d
\end{equation}
with the system size \cite{NatTang92}. As can be seen in the inset of 
Fig.~\ref{EWcoll} this really happens with the prefactor $c=0.30(1)$,
which agrees with the theoretical value $c= 2 a\simeq 0.304$ again. 

\begin{figure}[ht]
\begin{center}
\epsfxsize=70mm
\epsffile{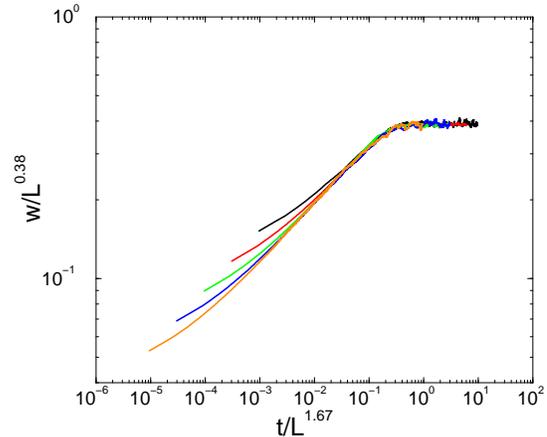}
\caption{(Color online) Scaling collapse for $p=1$, $q=0$ with
$2+1$ dimensional KPZ class exponents for different sizes: 
$L=64,128,256,512,1024$ (bottom to top).} \label{kpzcoll}
\end{center}
\end{figure}

For pure deposition $p=1$, $q=0$, or in case of other general up/down
asymmetric cases, we saw power-law increase of the surface width,
in agreement with the scaling hypothesis (\ref{FV-forf})
(see Fig.~\ref{kpzcoll}).
For the the largest system that we have investigated ($L=1024$) 
we fitted $W(t)$ in the $100 < t < 10000$ time window 
with a power-law and obtained $\beta=0.23(1)$. 
This value agrees quite well with the numerical estimate for
the $2+1$ dimensional KPZ class ($\beta=0.24$) provided in 
ref.~\cite{barabasi}. Note however that for smaller system sizes
the exponent estimate is somewhat smaller, due to corrections
to scaling, but one can clearly see a convergence towards 
higher values and a better collapse as $L\to\infty$.
Large scale simulations with an effective, bit-coded version of our 
algorithm could result in very precise estimates. The systematic 
tendency towards an asymptotic behavior has been found in 
Fig.~\ref{kpzcoll}.

The saturation values $W(inf.)$ for different system sizes 
also scale well with (\ref{FV-a}) and with the exponent
$\alpha=0.38(1)$ of the $2+1$ dimensional KPZ class \cite{MPP,barabasi}.
Assuming corrections to scaling of the form
$W\simeq A_2 L^{\alpha}(1 + B_2 L^{-\omega}$ the fitting to our
data resulted in very small effect: $\alpha=0.377(15)$, which marginally
overlaps with the value of \cite{MPP} but does not support the proposal
$\alpha=2/5$ of \cite{L98}.
Using these surface exponents and the scaling law (\ref{zlaw})
we estimated the dynamical exponent to be: $z=1.64(1)$,
which is somewhat greater than what one finds for the $2+1$ 
dimensional KPZ class in \cite{barabasi} ($z=1.58$). 
We think that this is due to the correction to scaling
observed in the time dependence discussed above.
If we scale the time with the dynamical exponent $z=1.64$ 
we obtain a good scaling collapse of the growth data 
for different sizes (Fig.\ref{kpzcoll})
in agreement with the (\ref{FV-forf},\ref{FV-a}) law again.
Our exponent estimates also satisfy the $\alpha+z=2$ scaling
relation within error margin. This implies that the Galilean 
invariance holds and the lattice model indeed lies in 
the $2+1$ dimensional KPZ universality class.

\section{Conclusions and outlook}

We have pointed out the possibility of mapping of a discrete 
surface growth processes onto a conserved, driven lattice gas 
model of oriented dimers, which move perpendicularly 
in two dimensions. The straight line motion of dimers 
in the two dimensional space is very similar to the motion 
of particles of the ASEP process. 
The difference is that since the dimers are extended objects, 
their motion is slowed down by the dimer particle 
exclusion and the sub-lattice update as compared to the single
particles of the ASEP. 
As a consequence their motion is described by somewhat 
larger dynamical exponent ($z \simeq 1.64$) than that of the 
ASEP ($z=3/2$), so the change of $z(d)$ seems to be a purely
topological phenomena in KPZ. This provides a better
understanding of the relation of universality classes 
of surface classes to those of the reaction-diffusion models. 
\cite{Orev,Obook08,dimerlcikk}. Interestingly the
$x/y$ symmetric surface dynamics maps onto a strongly
anisotropic reaction-diffusion model.

We have found KPZ or EW scaling by numerical simulations,
hence we showed that lattice anisotropy and under-surface
vacancies are irrelevant. Our simulation results for the 
$2+1$ dimensional EW case reproduced the theoretically expected
logarithmic scaling, with the correct leading order coefficients.
For the KPZ scaling our roughness exponent result is in the middle 
of the range obtained by various numerical exponent estimates: 
i.e. between $\alpha=0.36$ \cite{Ghai,Hasel} and the field 
theoretical value $\alpha=0.4$ \cite{L98}.
Our $\alpha=0.377(15)$ coincides with that of the numerical study 
\cite{AA04} and agrees with the renormalization results 
$\alpha=0.38$ \cite{CM01}. It overlaps marginally with the 
simulation results $\alpha=0.393(3)$ \cite{MPP} as well.
Our growth exponent estimate $\beta=0.23(1)$ matches the
results of \cite{Ghai} ($\beta=0.221(2)$) and \cite{AA04}
($\beta=0.229(5)$), obtained by independent numerical fitting
procedures. The dynamical exponent of this study is also in 
the range provided in \cite{AA04}.

Our model provides an efficient way of simulations and opens us
the possibility to study more complex growth models relevant
in recent interest of self-organizing surface nanosystem 
\cite{FacskoS}. An optimized, bit-coded version of our code, 
which manipulates the two-dimensional bit-field by logical 
operations runs roughly 10 times faster than the current version
and will be published elsewhere. 
For example the Bradley-Harper \cite{BH88} and the debated 
Kuramoto-Sivashinsky \cite{ks:1977} models with their modifications 
can be investigated numerically and are the subject of a 
forthcoming publications.

\vskip 1.0cm

\noindent
{\bf Acknowledgments:}\\

We thank Zoltan R\'acz and Uwe T\"auber for the useful comments.
Support from the Hungarian research fund OTKA (Grant No. T046129),
the bilateral German-Hungarian exchange program DAAD-M\"OB 
(Grant Nos. D/07/00302, 37-3/2008) and from the German Science 
Foundation (DFG research group 845, project HE2137/4-1) is 
acknowledged. The authors thank for the access to the HUNGRID.

\bibliography{ws-book9x6}

\end{document}